\documentclass[twocolumn,
superscriptaddress,
 amsmath,amssymb,
 aps,
]{revtex4-1}
\usepackage{graphicx}% Include figure files
\usepackage{dcolumn}% Align table columns on decimal point
\usepackage{bm}
\usepackage[version=3]{mhchem}
\usepackage{units}
\usepackage{upgreek}
\usepackage{braket}

\begin{document}

\newcommand{\dd}{\mathrm d}
\newcommand{\ii}{\mathrm i}
\newcommand{\muN}[1]{\mu_{\it N}^{#1}}
\newcommand{\efm}[1]{{\it e}^2\unit{fm}^{#1}}
\newcommand{\efmME}[1]{{\it e}\unit{fm}^{#1}}
\newcolumntype{L}[1]{>{\raggedright\arraybackslash}p{#1}} % linksbündig mit Breitenangabe
\newcolumntype{C}[1]{>{\centering\arraybackslash}p{#1}} % zentriert mit Breitenangabe
\newcolumntype{R}[1]{>{\raggedleft\arraybackslash}p{#1}} % rechtsbündig mit Breitenangabe

\preprint{APS/123-QED}

\title{Structure of high-lying levels populated in the $^{96}$Y $\rightarrow ^{96}$Zr $\beta$ decay}% Force line breaks with \\
%\thanks{A footnote to the article title}%

\author{K.~R.~Mashtakov}
%\homepage[\\Internet: ]{www.ikp.tu-darmstadt.de}
\affiliation{School of Computing, Engineering, and Physical Sciences, University of the West of Scotland, High Street, Paisley PA1 2BE, UK}
\affiliation{SUPA, Scottish Universities Physics Alliance, UK}

\author{V.~Yu.~Ponomarev}
\affiliation{Institut f\"ur Kernphysik, TU Darmstadt, Schlossgartenstr.~9, 64289 Darmstadt, Germany}

\author{M.~Scheck}
\email[Email: ]{marcus.scheck@uws.ac.uk}
\affiliation{School of Computing, Engineering, and Physical Sciences, University of the West of Scotland, High Street, Paisley PA1 2BE, UK}
\affiliation{SUPA, Scottish Universities Physics Alliance, UK}

\author{S.~W.~Finch}
\affiliation{Department of Physics, Duke University, Durham, NC 27708-0308, USA}
\affiliation{TUNL, Triangle Universities Nuclear Laboratory, Durham, NC 27708-0308, USA}

\author{J.~Isaak}
\affiliation{Institut f\"ur Kernphysik, TU Darmstadt, Schlossgartenstr.~9, 64289 Darmstadt, Germany}

\author{M.~Zweidinger}
\affiliation{Institut f\"ur Kernphysik, TU Darmstadt, Schlossgartenstr.~9, 64289 Darmstadt, Germany}

\author{O.~Agar}
\affiliation{Department of Physics, Karamanoglu Mehmetbey University, 70100 Karaman, Turkey}

\author{C.~Bathia}
\affiliation{Department of Physics, Duke University, Durham, NC 27708-0308, USA}
\affiliation{TUNL, Triangle Universities Nuclear Laboratory, Durham, NC 27708-0308, USA}

\author{T.~Beck}
\affiliation{Institut f\"ur Kernphysik, TU Darmstadt, Schlossgartenstr.~9, 64289 Darmstadt, Germany}

\author{J.~Beller}
\affiliation{Institut f\"ur Kernphysik, TU Darmstadt, Schlossgartenstr.~9, 64289 Darmstadt, Germany}

\author{M.~Bowry}
\affiliation{School of Computing, Engineering, and Physical Sciences, University of the West of Scotland, High Street, Paisley PA1 2BE, UK}
\affiliation{SUPA, Scottish Universities Physics Alliance, UK}

\author{R.~Chapman}
\affiliation{School of Computing, Engineering, and Physical Sciences, University of the West of Scotland, High Street, Paisley PA1 2BE, UK}
\affiliation{SUPA, Scottish Universities Physics Alliance, UK}

\author{M.~M.~R.~Chishti}
\affiliation{School of Computing, Engineering, and Physical Sciences, University of the West of Scotland, High Street, Paisley PA1 2BE, UK}
\affiliation{SUPA, Scottish Universities Physics Alliance, UK}

\author{U.~Friman-Gayer}
\affiliation{Institut f\"ur Kernphysik, TU Darmstadt, Schlossgartenstr.~9, 64289 Darmstadt, Germany}
\affiliation{TUNL, Triangle Universities Nuclear Laboratory, Durham, NC 27708-0308, USA}
\affiliation{Department of Physics and Astronomy, University of North Carolina at Chapel Hill, Chappel Hill, NC 27599, USA}

\author{L.~P.~Gaffney}
\affiliation{School of Computing, Engineering, and Physical Sciences, University of the West of Scotland, High Street, Paisley PA1 2BE, UK}
\affiliation{SUPA, Scottish Universities Physics Alliance, UK}

\author{P.~E.~Garrett}
\affiliation{Department of Physics, University of Guelph, Guelph, Ontario, Canada N1G 2W1}

\author{E.~T.~Gregor}
\affiliation{School of Computing, Engineering, and Physical Sciences, University of the West of Scotland, High Street, Paisley PA1 2BE, UK}
\affiliation{SUPA, Scottish Universities Physics Alliance, UK}

\author{J.~M.~Keatings}
\affiliation{School of Computing, Engineering, and Physical Sciences, University of the West of Scotland, High Street, Paisley PA1 2BE, UK}
\affiliation{SUPA, Scottish Universities Physics Alliance, UK}

\author{U.~K\"oster}
\affiliation{ILL, Institut Laue-Langevin, 71 avenue des Martyrs, 38000 Grenoble, France}

\author{B.~L\"oher}
\affiliation{GSI Helmholtzzentrum f\"ur Schwerionenforschung GmbH, Darmstadt, 64289 Darmstadt, Germany}

\author{A.~D.~MacLean}
\affiliation{Department of Physics, University of Guelph, Guelph, Ontario, Canada N1G 2W1}

\author{D.~O'Donnell}
\affiliation{School of Computing, Engineering, and Physical Sciences, University of the West of Scotland, High Street, Paisley PA1 2BE, UK}
\affiliation{SUPA, Scottish Universities Physics Alliance, UK}

\author{H.~Pai}
\affiliation{Institut f\"ur Kernphysik, TU Darmstadt, Schlossgartenstr.~9, 64289 Darmstadt, Germany}
\affiliation{Nuclear Physics Division, Saha Institute of Nuclear Physics, Kolkata-700064, India}

\author{N.~Pietralla}
\affiliation{Institut f\"ur Kernphysik, TU Darmstadt, Schlossgartenstr.~9, 64289 Darmstadt, Germany}

\author{G.~Rainovski}
\affiliation{Faculty of Physics, University of Sofia St.~Kliment Ohridski, 1164 Sofia, Bulgaria}

\author{M.~Ramdhane}
\affiliation{LPSC, UJF Grenoble I, 53 avenue des Martyrs, 38026 Grenoble Cedex, France}

\author{C.~Romig} 
\affiliation{Institut f\"ur Kernphysik, TU Darmstadt, Schlossgartenstr.~9, 64289 Darmstadt, Germany}

\author{G.~Rusev}
\affiliation{Department of Physics, Duke University, Durham, NC 27708-0308, USA}
\affiliation{TUNL, Triangle Universities Nuclear Laboratory, Durham, NC 27708-0308, USA}

\author{D.~Savran}
\affiliation{GSI Helmholtzzentrum f\"ur Schwerionenforschung GmbH, Darmstadt, 64289 Darmstadt, Germany}

\author{G.~S.~Simpson}
\affiliation{LPSC, UJF Grenoble I, 53 avenue des Martyrs, 38026 Grenoble Cedex, France}

\author{J.~Sinclair}
\affiliation{School of Computing, Engineering, and Physical Sciences, University of the West of Scotland, High Street, Paisley PA1 2BE, UK}
\affiliation{SUPA, Scottish Universities Physics Alliance, UK}

%\author{J.~F.~Smith}
%\affiliation{School of Computing, Engineering, and Physical Sciences, University of the West of Scotland, High Street, Paisley PA1 2BE, UK}
%\affiliation{SUPA, Scottish Universities Physics Alliance, UK}

\author{K.~Sonnabend}
\affiliation{ILL, Institut Laue-Langevin, 71 avenue des Martyrs, 38000 Grenoble, France}

\author{P.~Spagnoletti}
\affiliation{School of Computing, Engineering, and Physical Sciences, University of the West of Scotland, High Street, Paisley PA1 2BE, UK}
\affiliation{SUPA, Scottish Universities Physics Alliance, UK}

\author{A.~P.~Tonchev}
\affiliation{Department of Physics, Duke University, Durham, NC 27708-0308, USA}
\affiliation{TUNL, Triangle Universities Nuclear Laboratory, Durham, NC 27708-0308, USA}

\author{W.~Tornow}
\affiliation{Department of Physics, Duke University, Durham, NC 27708-0308, USA}
\affiliation{TUNL, Triangle Universities Nuclear Laboratory, Durham, NC 27708-0308, USA}

%\author{V.~Werner}
%\affiliation{Institut f\"ur Kernphysik, TU Darmstadt, Schlossgartenstr.~9, 64289 Darmstadt, Germany}

% \collaboration{The EXILL Nd-144-Collaboration}

\date{\today}% It is always \today, today,
             %  but any date may be explicitly specified

\begin{abstract}
 {The nature of $J^{\pi}=1^-$ levels of $^{96}$Zr below the $\beta$-decay Q value of $^{96}$Y has been investigated in high-resolution $\gamma$-ray spectroscopy following the $\beta$ decay as well as in a campaign of inelastic photon scattering experiments. Branching ratios extracted from $\beta$ decay allow the absolute $E1$ excitation strength to be determined for levels populated in both reactions. The combined data represents a comprehensive approach to the wavefunction of $1^-$ levels below the $Q_{\beta}$ value, which are investigated in the theoretical approach of the Quasiparticle Phonon Model. This study clarifies the nuclear structure properties associated with the enhanced population of high-lying levels in the $^{96}$Y$_{gs}$ $\beta$ decay, one of the three most important contributors to the high-energy reactor antineutrino spectrum.}
  
%  In $^{144}$Nd the $3_3^-$ state at $2778\,\mathrm{keV}$ is a good candidate for such a ``mixed-symmetry'' octupole state. In order to clarify the nature of this state, a $^{143}\mathrm{Nd}(\mathrm{n}, \gamma)$-experiment was conducted with the EXILL-setup. Following neutron capture the $3^-$ states are populated and EXILL provides the opportunity to determine the multipole-mixing ratios of the $3_i^-\rightarrow 3_1^-$ transitions. For a transition from the isovector one-phonon state to the isoscalar one phonon state a large $M1$ component in the order of $\unit[1]{\muN{}}$, and thus a small multipole-mixing ratio, is predicted.
  
%  As a second part of the project the lifetime of the $3_3^-$ state was measured with GAMS6, a high-resolution spectrometer to mesure lifetimes in the femtosecond region. Both values, the multipole-mixing ratio and the lifetime, lead to an absolute transtion strength helping us to understand the nature of the $3_i^-$ states and the proton-neutron degree of freedom in the octupole sector.
%\begin{description}
%    \item[Usage] Secondary publications and information retrieval purposes.
%   \item[PACS numbers] %May be entered using the \verb+\pacs{#1}+ command.
%    \item[Structure] You may use the \texttt{description} environment to structure your abstract; use the optional argument of the \verb+\item+ command to give the category of each item. 
 % \end{description}
\end{abstract}

%\pacs{Valid PACS appear here}% PACS, the Physics and Astronomy
                             % Classification Scheme.
%\keywords{Suggested keywords}%Use showkeys class option if keyword
%                                display desired
\maketitle
Following the observation of a lack of high-energy antineutrinos emitted from nuclear reactors \cite{An}, nuclear \mbox{$\beta$-decay} studies using total absorption $\gamma$-ray spectroscopy (TAGS) have shown that a larger number of $\beta$ decays populate high-lying excited levels of the daughter nucleus (e.g., see Refs.~\cite{RascoI,Tain} and references therein) than hitherto believed. Hence, the average energy shared by the emitted electron and antineutrino is less than previously anticipated from $\beta$-decay studies using high resolution $\gamma$-ray spectroscopy (HRS). While antineutrinos escape the reactor without further interaction, $\gamma$ rays or conversion electrons emitted in the decay of these high-lying levels will be absorbed in the reactor and contribute to the heat production. Furthermore, the presence of high-energy $\gamma$ rays leads to the need for appropriate shielding for future more compact reactor designs. 

The underdetermined population of high-lying levels in previous $\beta$-decay studies can be attributed to the pandemonium effect \cite{Hardy}; that is, $\gamma$ rays which are emitted in a $\gamma$-$\gamma$ cascade are not recognised as such and are placed in the level scheme at too low energy or $\gamma$ rays are not observed at all. A major source of this underdetermination is related to the detectors that have been previously used. Many of the earlier relevant $\beta$-decay studies used first generation semiconductor detectors with low $\gamma$-ray detection efficiency in comparison to what is currently available. This is particularly true for $\gamma$~rays of comparatively high energy of several MeV. Meanwhile, TAGS has demonstrated the presence of this effect and quantified it to a good degree (e.g., see Ref.~\cite{Estienne}). However, a task that TAGS, with its highly-efficient scintillator detectors, but limited energy resolution cannot fulfill is to clarify the nature of the populated levels. Therefore, HRS is necessary to identify the populated levels and to determine their spectroscopic properties. In order to explore the nature of these levels, it is also beneficial to investigate them with complementary reactions; however, due to the high level density above about 3~MeV, this is a difficult task. For $\beta$ decays originating from mother nuclei with low ground-state spin, an ideal combination of population processes was identified in Ref.~\cite{Scheck}. There, $\beta$ decay was used to populate levels associated with the Pygmy Dipole Resonance (PDR) \cite{Deniz,AngelaI}, an accumulation of strongly excited $1^-$ levels on the low-energy tail of the Giant Dipole Resonance (GDR) \cite{AngelaII,Mushin}. A standard tool to investigate these $1^-$ levels in stable nuclei is the resonant scattering of real photons, the so-called nuclear resonance fluorescence (NRF) \cite{Kneissl}. Because of the low associated angular momentum transfer, which is almost entirely limited to the 1$\hbar$ intrinsic angular momentum of the photon, the $(\gamma,\gamma^{\prime})$ reaction selectively populates $1^{\pi}$ levels and permits $\gamma$-ray spectroscopy in energy regimes with a high level density. Of great advantage is that the NRF scattering process is solely governed by the well-understood electromagnetic interaction. Hence, in addition to the complex $\beta$-decay matrix element connecting the mothers ground state and the excited level in the daughter, for many levels the wavefunction is tested with the electromagnetic matrix element connecting the ground state of the daughter and the excited level. 

Interestingly, the three main contributors \cite{Libby} to the reactor high-energy antineutrino spectrum, $^{92}$Rb ($Q_{\beta} = 8095(6)$~keV) \cite{NDS113}, $^{96}$Y  ($Q_{\beta} = 7096(23)$~keV) \cite{NDS109}, and $^{142}$Cs ($Q_{\beta} = 7308(11)$~keV) \cite{NDS112} all have a $0^-$ ground state. Consequently, it can be expected that, in the daughter, mainly $1^-$ levels are populated via Gamow-Teller $\beta$ decays, exactly the type of levels that are populated in the NRF of their even-even daughter nuclei. However, the NRF cross sections are comparatively low and given the available photon flux at current facilities, the target material required limits the applicability to (quasi-)stable isotopes. For this reason, here the $\beta$ decay of $^{96}$Y to its stable daughter $^{96}$Zr was investigated,  and the combined results have been interpreted within the microscopic approach of the Quasi-particle Phonon Model (QPM) \cite{Soloviev}.% An additional benefit of $^{96}$Zr is that recent work dedicated to the low-energy level scheme \cite{Kramer} has identified a distinct difference in the microscopic structure of the $0^+$ ground state (almost no contribution of the proton ($\pi$) $\pi 1g_{9/2^+}$ and neutron ($\nu$) $\nu 1g_{7/2^+}$ subshells) and the first excited level at 1581~keV, which is a shape coexisting $0^+$ level with strong $\pi 1g_{9/2^+}$ and $\nu 1g_{7/2^+}$ contributions in its wavefunction. 

\begin{figure}
\includegraphics[width=8.0cm]{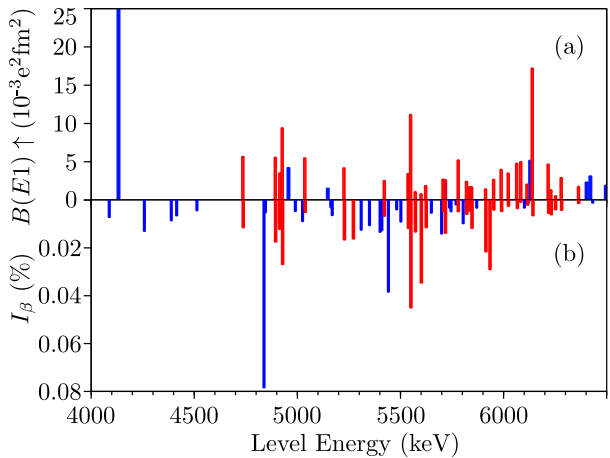}
\caption{\label{Figure2} (Color online) Comparison of the $^{96}$Zr $B(E1)$-strength distribution observed in the NRF reaction [part~(a)] and the population-intensity distribution in $\beta$ decay from $^{96}$Y$_{gs}$ [part~(b)]. Levels that are populated in both reactions are given as red bars and levels that are solely observed in one reaction as blue bars. The results from NRF were corrected for the branching ratios observed in $\beta$ decay.}
\end{figure}

The $\beta$-decay experiment was performed following the neutron-induced fission of $^{235}$U at the research reactor of the Institut Laue Langevin (ILL). Fission fragments were mass separated using the LOHENGRIN separator \cite{Lohengrin} and transported to a setup consisting of a cooled lithium-doped silicon detector used for the detection of electrons, two high efficiency Clover (HPGe) germanium detectors in close geometry and two further single-crystal HPGe detectors. This approach allowed the measurement of electron-$\gamma$ and $\gamma$--$\gamma$ coincidences. The extracted data will benefit the TAGS measurement for this particular decay, since the TAGS measurement was not sensitive to the 1581-keV $0^+_2 \rightarrow 0^+_{gs}$ $E0$ decay \cite{RascoII} from the first excited level in $^{96}$Zr. The mass separation of LOHENGRIN is not sufficient to distinguish between the $0^-$ ground state and the $8^+$ isomeric state of the $^{96}$Y mother. For levels in $^{96}$Zr below 4~MeV excitation energy it was necessary to rely on the $\beta$-decay data from Ref.~\cite{Henrik}, which does distinguish decays from the $0^-$ and $8^+$ states. In addition to available spectroscopic information, the distinct decay behaviour of levels populated by these very different spins allows an association of $\gamma$ rays originating from levels above 4~MeV with the $^{96}$Y ground-state decay. The data from Ref.~\cite{Henrik} were also used for an internal efficiency calibration for the present data. The improved overall $\gamma$-ray detection efficiency resulted in the identification of the population of additional levels and several newly-observed branching transitions to lower-lying levels. 

The $^{96}$Zr($\gamma,\gamma^{\prime}$) NRF campaign used continuous bremstrahlung beams at the DHIPS setup \cite{Kerstin} of the S-DALINAC as well as quasimonochromatic ($\Delta E_{\gamma}/E_{\gamma} \approx 3$~\%) fully-polarised photon beams at the High Intensity $\gamma$-ray Source (HI$\vec{\gamma}$S) \cite{Henry} in the entrance channel. From this campaign the spin and parity of the NRF excited levels were unambiguously determined and their integrated scattering cross sections measured. More details about the experiments, data analysis, and the final data tables will be reported in a following publication \cite{Finch}.

\begin{figure}
\includegraphics[width=8.0cm]{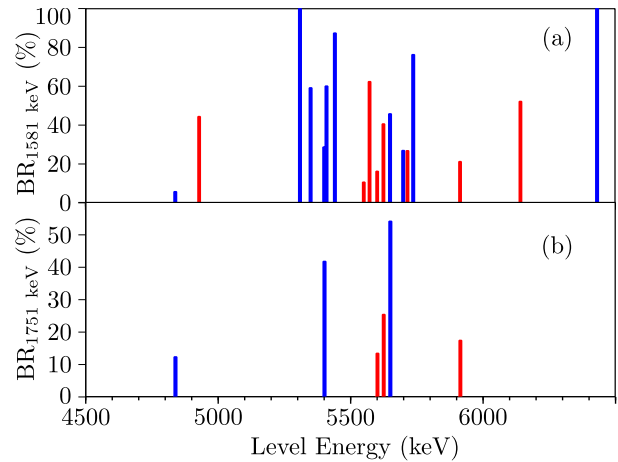}
\caption{\label{Figure3} (Color online) Branching ratios to the first excited $0^+_2$ level at 1581~keV [part~(a)], and the first excited $2^+_1$ level at 1751~keV [part~(b)] as extracted from the present $\beta$-decay measurement. Levels observed in both NRF and $\beta$ decay are marked with red bars. Levels that were solely observed in $\beta$~decay are marked as blue bars. Note the different scales on the y axis.}
\end{figure}

In Fig.~\ref{Figure2}, the level population in $\beta$ decay [part~(b)] is compared to the excitation strength of levels observed in NRF [part~(a)] up to 6.5~MeV. From the part of the NRF campaign that used fully-polarised $\gamma$ rays in the entrance channel, for all levels above 4.5~MeV excited in this reaction, a firm spin and parity assignment of $1^-$ has been made. This observation agrees with the expectation that most levels observed in $\beta$~decay are populated in Gamow-Teller allowed decays from the $0^-$ ground state of $^{96}$Y. Almost all levels observed in NRF were also observed in $\beta$ decay. Due to the large background in NRF spectra stemming from non-resonant scattered photons, which grows exponentially to lower energies, branching transitions are often below the sensitivity limit (Fig.~\ref{Figure3} shows the branching ratios for the two low-lying excited levels that are most frequently populated). Consequently, the excitation probabilities in the NRF data are underestimated. The $B(E1)$ excitation strengths shown in Fig.~\ref{Figure2} include the branching ratio data from $\beta$ decay. Hence, apart from a few levels with a $B(E1)$ strength below the sensitivity limit, the data presented in Fig.~\ref{Figure2}~(a) resembles the true excitation pattern up to 6.4~MeV.

\begin{figure}
\includegraphics[width=8.0cm]{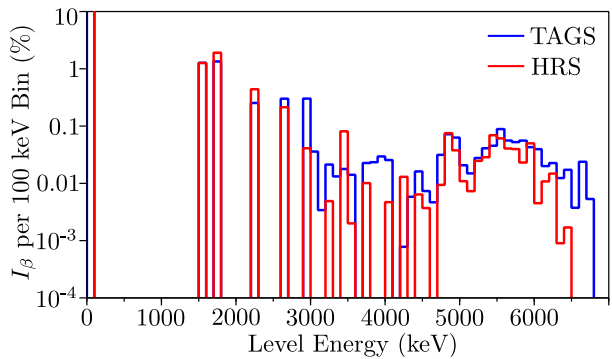}
\caption{\label{Figure1} (Color online) Comparison of level population intensity as a function of excitation energy between the results from total absorption $\gamma$-ray spectroscopy \cite{RascoII} (TAGS, blue) and the combination of the presented work and data published in Ref.~\cite{Henrik} for high-resolution $\gamma$-ray spectroscopy (HRS, red). For a discussion see text.}
\end{figure}

The level population probability in $\beta$ decay extracted from the present HRS measurement (in combination with the data from Ref.~\cite{Henrik}) and the data obtained in the TAGS work \cite{RascoII} are compared in Fig.~\ref{Figure1}. In order to be able to compare the results, the data of HRS and TAGS are given in 100~keV bins. Remarkably, both methods result in overall good agreement. The two most striking differences are groups of levels between 3 and 4~MeV in the TAGS spectrum and the very high-lying levels that were not seen in HRS. As shown in Fig.~\ref{Figure3}~(a) in HRS several levels, especially a group of levels near 5.5~MeV, strongly branch to the first excited $0^+$ level at 1581~keV. Since the TAGS measurement reported in Refs.~\cite{RascoI,RascoII} was not sensitive to the subsequent $E0$ transition, these $\gamma$ rays were placed at a too low energy. The non-observation of the very high-lying levels in HRS has sensitivity issues. The summed singles $\gamma$-ray spectrum from all HPGe detectors was contaminated with a large background stemming from the nearby reactor. In order to suppress the uncorrelated background, this work used $e^-$--$\gamma$ coincidences. However, the threshold of the silicon electron detector was set at a comparably high energy of $\approx$300~keV and, consequently, a large fraction of the $\beta$-particle spectrum for decays to levels near the $Q_\beta$ value lies below the energy threshold and was consequently not included. Hence, the $e^-$--$\gamma$ coincidence efficiency of these weakly populated levels was too low for them to be observed. Nevertheless, Fig.~\ref{Figure1} demonstrates that HRS using modern highly-efficient arrays is capable of resolving the pandemonium effect and is able to establish the $1^-$ levels of the PDR as final states of the $^{96}$Y $\beta$ decay.

\begin{figure}
\includegraphics[width=8.6cm]{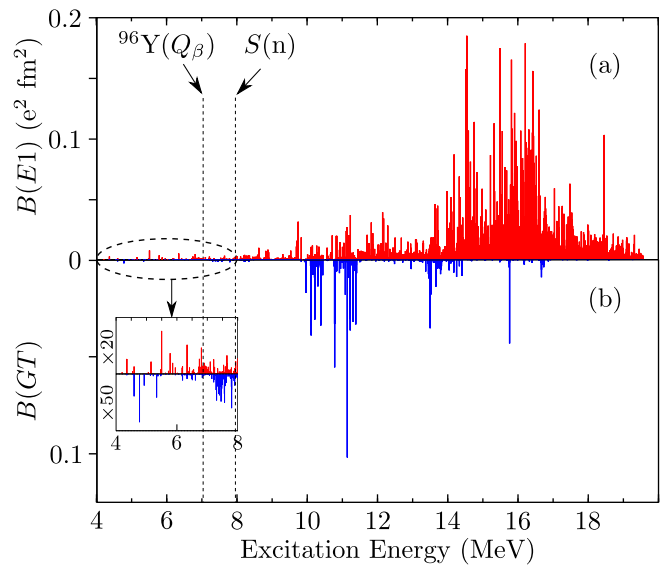}
\caption{\label{f1}  (Color online) The upper part shows the $B(E1)$-strength distribution of $^{96}$Zr as calculated in the quasiparticle phonon model. In the lower part the calculated $B(GT)$-strength distribution for the $^{96}$Y $\rightarrow ^{96}$Zr $\beta$ decay is shown. The experimentally observable $B(GT)$ strength below $Q_\beta \approx 7.1$~MeV is shown in the inset. For a discussion see text.}
\end{figure}

In the following, the properties of the $1^-$ states in $^{96}$Zr are investigated in the QPM \cite{Soloviev,IP12}. The model calculations have been performed with wavefunctions that contain one-, two-, and three-phonon configurations. For computational reasons, the configurations above 25.0, 21.0, and 9.7~MeV, respectively, have been truncated. This truncation allows a consideration of the fragmentation of the excitation strength in a wide energy interval including the GDR and at the same time allows a study of the fine structure of the PDR in more detail. In order to accommodate Gamow-Teller (GT) decays, the model approach outlined in Ref.~\cite{Scheck}, which exclusively contained Fermi decays, had to be significantly modified. 

The calculated $B(E1)$-strength distribution in $^{96}$Zr is presented in Fig.~\ref{f1}~(a). It is dominated by the GDR with the energy centroid at 16.3~MeV. Fig.~\ref{f1}~(b) provides the $B(GT)$ transition strength from the $0^-$ ground state of $^{96}$Y to the same set of $1^-$ states in $^{96}$Zr. In this calculation, we assume the
wavefunction of the $^{96}$Y ground state as a pure $\{\nu 3s_{1/2} \times \pi 2p_{1/2}\}_{0^-}$ configuration. This assumption is confirmed by a microscopic QRPA calculation, which yields more than a 99$\%$ contribution of this configuration to the wavefunction of the first $0^-$ state in $^{96}$Y. Interestingly, in order to couple to $J=0$, the spins of proton and neutron must align to $S=1$. Approximately 96~\% of all $\beta$ decays \cite{RascoI,Henrik} proceed to the ground state of $^{96}$Zr. In this first-forbidden decay \cite{Suhon}, the neutron in the $\nu 3s_{1/2}$ subshell decays with a spin-flip to the proton $\pi 2p_{1/2}$ subshell. The $B(GT)$ strength [Fig.~\ref{f1}~(b)] is concentrated at much lower energies as compared to the $B(E1)$ strength [Fig.~\ref{f1}~(a)] with only weak $B(GT)$ transitions at the GDR peak.

\begin{figure}
\includegraphics[width=8.6cm]{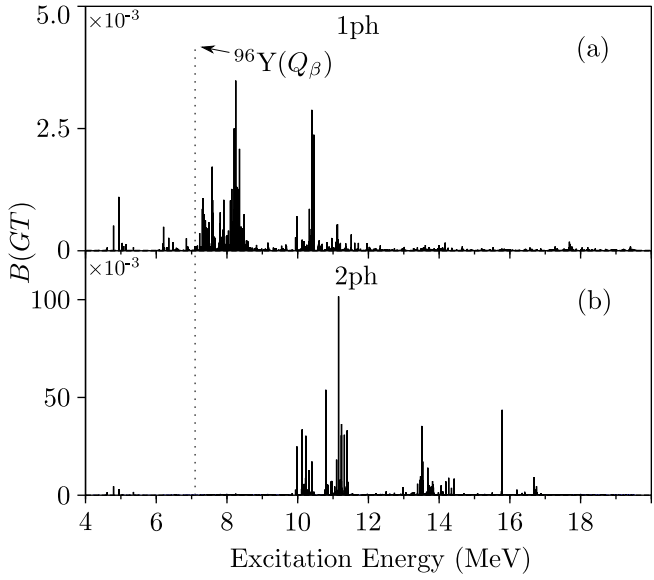}
\caption{\label{f2} $B(GT)$ Gamow Teller-strength distribution of the $^{96}$Y $\rightarrow ^{96}$Zr $\beta$ decay disentangled whether one-phonon (1p1h) [part~a)] or two-phonon (2p2h) [part~b)] components are populated in the wavefunction of the final $1^-$ level. For a clarity of the presentation the scales on the y-axis were adjusted. The total $B(GT)$ strength including interference between the various components is shown in Fig.~\ref{f1}~(b).}
\end{figure}

There are other essential differences in the generation of the $B(E1)$ and $B(GT)$ strengths. As already discussed in Refs.~\cite{Scheck,Her97,Pon98}, excited states of even-even nuclei are predominantly populated in the electromagnetic excitation from the ground state via one-phonon components of their wavefunctions; two-phonon configurations are excited much more weakly (except for the collective $[2^+_1\otimes 3^-_1]_{1^-}$ state) and form a kind of a structureless ``background''. In contrast, the $B(GT)$ matrix elements for transitions to two-phonon configurations dominate over transitions to one-phonon configurations. Fig.~\ref{f1}~(b) resembles Fig.~\ref{f2}~(b), in which transitions to two-phonon components of the wavefunction of 1$^-$ states are plotted. $B(GT)$ transitions to one-phonon components are presented in the top part of Fig.~\ref{f2}; they are much weaker than the former ones (compare the scales in Fig.~\ref{f2}) and are located at lower energies.

\begin{figure}
\includegraphics[width=8.6cm]{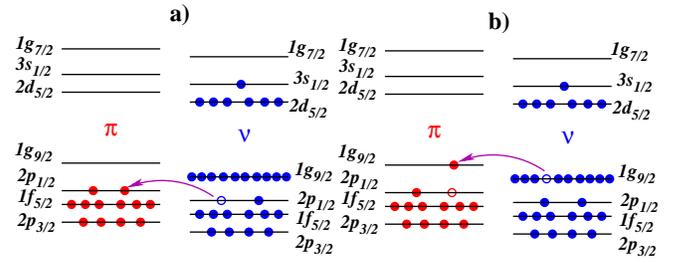}
\caption{\label{Figure4} (Color online) Schematic representation of the shell structure involved in the $\beta$ decay of $^{96}$Y. Part~a) shows the decay to the neutron $\{2p_{1/2}^{-1} \times 3s_{1/2}\}_{\nu}$ one-particle one-hole configuration and part~b) the decay resulting in a proton $\{2p_{1/2}^{-1} \times 1g_{9/2}\}_{\pi}$ and neutron $\{1g_{9/2}^{-1} \times 3s_{1/2}\}_{\nu}$ two-particle two-hole configuration. Particles are shown as full circles, while holes are represented by open circles.}
\end{figure}

In the QPM, phonons are composed of proton and neutron one-particle one-hole (1p1h) configurations. Their excitation energies and corresponding internal fermionic structure is obtained from solving the quasi-particle random phase approximation equations. While many 1p1h configurations have a non-zero matrix element for E$\lambda$ transitions and, accordingly, contribute to the $B(E\lambda)$ value for one-phonon components of the wavefunction, the population of one-phonon components in the $\beta$-decay of odd-odd nuclei is very selective and governed by only a few 1p1h configurations. For the $^{96}$Y$\to^{96}$Zr Gamow-Teller decay the relevant configurations are: $\{2p_{1/2}^{-1} \times 3s_{1/2}\}_{\pi}$ at 10.1~MeV, $\{2p_{1/2}^{-1} \times 3s_{1/2}\}_{\nu}$ at 8.8~MeV, and $\{2p_{3/2}^{-1} \times 3s_{1/2}\}_{\nu}$ at 10.5~MeV, which correspond to the decays $3s_{1/2} \to 3s_{1/2}$, $2p_{1/2} \to 2p_{1/2}$, and $2p_{3/2} \to 2p_{1/2}$, respectively. A visualisation for a decay leaving the daughter in the $\{2p_{1/2}^{-1} \times 3s_{1/2}\}_{\nu}$ 1p1h configuration is shown in Fig.\ref{Figure4}~a). The residual interaction of the nuclear Hamiltonian mixes these configurations in the set of one-phonon $1^-$ states and coupling to complex two- and three-phonon configuration leads to further damping of the GT strength [see Fig.~\ref{f2}~(a)]. An example for the population of two-phonon components, corresponding to a two-particle two-hole (2p2h) excitation, is shown in Fig.~\ref{Figure4}~b).

According to the QPM predictions, only a small fraction of the $B(GT)$ strength may be observed in the \mbox{$\beta$-decay} experiment for energy reasons (see, dotted line in Fig.~\ref{f1} showing the $Q_{\beta}$ value). At the same time, the model provides a reasonable fragmentation of the Gamow-Teller strength and the absolute $B(GT)$ values at low energies
(see inset in Fig.~\ref{f1}). As a reference, we compare the most strongly populated $1^-$ level at 4.838~MeV with an experimental $\log ft = 6.64$ to the calculated state with the strongest population at 4.895~MeV with $\log ft = 6.66$ (calculated with $g_A/g_V = 1.23$ \cite{BM1}). Furthermore, the model predicts a rather complex wavefunction, which is dominated by configurations that are not populated in $\beta$ decay. The $\{2d_{5/2}^{-1} \times 3s_{1/2}\}_{\nu}\{2p_{1/2}^{-1} \times 2d_{5/2}\}_{\pi}$ 2p2h component, which is populated in the $\nu2d_{5/2}\rightarrow \pi2d_{5/2}$ $\beta$~decay, contributes only with 0.27~\% to the normalisation of the wavefunction of 4.77~MeV state. Experimentally, the complexity of the wavefunction of this level is evident by the observed eight $\gamma$-ray decays to lower-lying states.

%Levels that are strongly populated in both reactions can be identified to have a noticable $[\nu 3s_{1/2},(\nu 2p_{1/2})^{-1}]$ component in their wavefunction. Examples are the three levels just below 5.0~MeV and the two levels just above 5.5~MeV. However, the majority of $\beta$ decays to excited levels can be expected to leave the daughter in a 2p2h configuration (see Fig.~\ref{Figure4}~b). Specifically these are the decays of the neutron in the $\nu 2d_{5/2}$ and $\nu 1g_{9/2}$ subshells. One condition for the identification of levels with a considerable 2p2h component in their wavefunction is a weak or even vanishing $B(E1)$ excitation probability. Furthermore, the 2p2h configurations are expected to decay to 1p1h components in lower-lying excited states. Consequently, the in NRF weakly or not populated levels near 5.5~MeV, with their strong decay branches to lower-lying excited states, are expected to be populated via their 2p2h components.

\begin{figure}
\includegraphics[width=8.6cm]{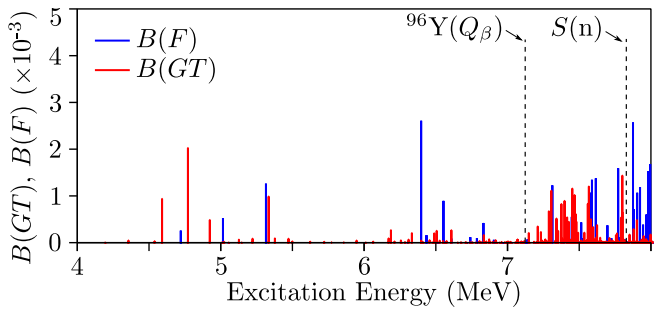}
\caption{\label{f3}  (Color online) Calculated low-energy part of the $\beta$-decay strength of the $^{96}$Y $J^{\pi}=0^-$ ground-state decay. The red bars shows the Gamow-Teller $B(GT)$ strength populating $1^-$ levels and the blue bars the Fermi $B(F)$ strength populating $0^-$ levels in $^{96}$Zr. The experimental $Q_{\beta}$ value and neutron separation energy $S_n$ are indicated by the dashed lines.}
\end{figure}

For completeness, we have also calculated $\beta$-decay of $^{96}$Y to $0^-$ states of $^{96}$Zr by Fermi transitions. The $0^-$ states are described by wavefunctions of the same complexity as for the $1^-$ states. %According to theoretical predictions some of the levels observed in $\beta$-decay and not populated in the NRF experiment may be $0^-$ and not $1^-$ states. 
While the calculation predicts several $0^-$ levels below $Q_{\beta}$ (see Fig.~\ref{f3}), a search in the $\gamma$-$\gamma$ matrix for $E1$/$M1$ decays to low-lying $1^+$/$1^-$ levels did not reveal any experimental candidates for $0^-$ levels.

In this contribution the nature of $1^-$ levels in $^{96}$Zr is investigated in a multi-messenger approach using the $^{96}$Y ($J^{\pi}=0^-$) $\beta$ decay and the $^{96}$Zr($\gamma,\gamma^{\prime}$) reaction, which is meanwhile a standard approach \cite{Deniz19}. However, usually a combination of inelastic scattering experiments, which are mostly sensitive to the 1p1h components, is used. The NRF-$\beta$ decay approach employed in this work provides the strongest relative difference in the population of the excited levels and, consequently, represents a very sensitive probe of the wavefunctions. A comparison to TAGS data reveals that HRS employing modern HPGe detectors is capable of resolving the pandemonium effect. Furthermore, in the energy range where the present $\beta$-decay measurement is sensitive, basically all $1^-$ levels that are excited in the $(\gamma,\gamma^{\prime})$ reaction are populated. This observation demonstrates the role of the PDR in the population of higher-lying levels in the $\beta$ decay of $^{96}$Y, which is an important contributor to the reactor antineutrino anomaly. The investigation of the microscopic structure of the observed levels resulted in an excellent agreement, clarifying which components in the complex wavefunctions of the PDR $1^-$ levels are populated. 

This work is dedicated to the memory of Henryk Mach. The authors acknowledge the generous offer by B.~C.~Rasco to provide the data for the TAGS measurement used in Figure~1. The UK authors acknowledge financial support by the UK-STFC. This work is supported by the Deutsche Forschungsgemeinschaft through Grant~No.~\mbox{SFB-634 (Project ID 5465852)} and \mbox{SFB-1245 (279384907)} and by the U.S. Department of Energy (DOE), Office of Nuclear Physics, under grant No.~DE-FG02-97ER41033. H.P. is grateful for the support of the Ramanujan Fellowship research grant under SERB-DST (SB/S2/RJN-031/2016).  

\bibliography{Nd144}

\begin{thebibliography}{99}

\bibitem{An} F.~P.~An et al., Phys.~Rev.~Lett.~{\bf 116}, 061801 (2016).
\bibitem{RascoI} B.~C.~Rasco et al., Phys.~Rev.~Lett.~{\bf 117}, 092501 (2016).
\bibitem{Tain} V.~Guadilla et al., Phys.~Rev.~Lett.~{\bf 122}, 042502 (2019).  
\bibitem{Hardy} J.~C.~Hardy, L.~C.~Carraz, B.~Jonson, and P.~G.~Hansen,
Phys.~Lett.~{\bf 71B}, 307 (1977).
\bibitem{Estienne} M.~Estienne, et al., Phys.~Rev.~Lett.~{\bf 123}, 022502 (2019).
\bibitem{Scheck} M.~Scheck et al., Phys.~Rev.~Lett.~{\bf 116}, 132501 (2016). 
\bibitem{Deniz} D.~Savran et al., Prog.~Part.~Nucl.~Phys.~{\bf 70}, 210 (2013).
\bibitem{AngelaI} A.~Bracco, E.~G.~Lanz, and A.~Tamii, Prog.~Part. Nucl.~Phys. {\bf 106}, 360 (2019).
\bibitem{AngelaII} P.~F.~Bortignon, A.~Bracco, R.~A.~Broglia, {\it Giant Resonances}, (CRC Press, 1998). 
\bibitem{Mushin} M.~Harakeh and A.~van~der~Woude, {\it Giant Resonances} (Oxford University Press, Oxford, 2001).
\bibitem{Kneissl} U.~Kneissl, H.~H.~Pitz, and A.~Zilges, Prog.~Part. Nucl.~Phys.~{\bf 37}, 349 (1996).  
\bibitem{Libby} A.~A.~Sonzogni, T.~D.~Johnson, and E.~A.~McCutchan, Phys.~Rev.~C {\bf 91}, 011301 (2015).
\bibitem{NDS113} C.~M.~Baglin,  Nucl.~Data~Sheets {\bf 113}, 2187 (2012).
\bibitem{NDS109} D.~Abriola and A.~A.~Sonzogni, Nucl.~Data~Sheets {\bf 109}, 2501 (2008).
\bibitem{NDS112} T.~D.~Johnston, D.~Symochko, M.~Fadil, and J.~K.~Tuli, Nucl.~Data~Sheets {\bf 112}, 1949 (2011).
\bibitem{Soloviev} V.~G.~Soloviev, {\it  Theory of Atomic Nuclei, Quasiparticles and Phonons} (IOP, London, 1992).
%\bibitem{Kramer} C.~Kremer et al., Phys.~Rev.~Lett.~{\bf 117}, 172503 (2016).
\bibitem{Lohengrin} P.~Armbruster et al., Nucl.~Inst.~Methods Phys. Res. {\bf 139}, 213 (1976). 
\bibitem{RascoII} B.~C.~Rasco et al., Acta.~Phys.~Pol.~{\bf 48} (2017) 507.
\bibitem{Henrik} H.~Mach et al., Phys.~Rev.~C {\bf 41} (1990) 226.
\bibitem{Kerstin} K.~Sonnabend et al., Nucl.~Instrum.~Methods Phys. Res.~A ~{\bf 640}, 6 (2011). 
\bibitem{Henry} H.~W.~Weller et al., Prog.~Part.~Nucl.~Phys.~{\bf 62}, 257 (2009).
\bibitem{Finch} S.~W.~Finch, J.~Isaak, K.~R.~Mashtakov et al., in preparation
\bibitem{Suhon} L.~Hayen, J.~Kostensalo, N.~Severijns, and J.~Suhonen, Phys.~Rev.~C {\bf 100}, 054323 (2019)
\bibitem{IP12} N.~Lo~Iudice, V.~Yu.~Ponomarev, Ch.~Stoyanov, A.~V.~Sushkov, and V.~V.~Voronov, J.~Phys.~G {\bf 39}, 043101 (2012).
\bibitem{Her97} R.-D.~Herzberg {\it at al.}, 
%P. von Brentano, J. Eberth, J. Enders, R. Fischer, N. Huxel, T. Klemme, 
%P. von Neumann-Cosel, N. Nicolay, N. Pietralla, V.Yu. Ponomarev, 
%J. Reif, A. Richter, C. Schlegel, R. Schwengner, S. Skoda, H. G. Thomas, 
%I. Wiedenhover, G. Winter, A. Zilges
Phys.~Lett.~B {\bf 390}, 49 (1997). 
\bibitem{Pon98}
V.~Yu.~Ponomarev, Ch.~Stoyanov, N.~Tsoneva, and M.~Grinberg, Nucl. Phys. A {\bf 635}, 470 (1998).
\bibitem{BM1} A.~Bohr and B.~Mottelson, {\it  Nuclear Structure, vol. 1: 
Single-Particle Motion}, (Benjamin, New York, 1969). 
\bibitem{Deniz19} D.~Savran et al., Phys.~Lett.~B {\bf 786}, 16 (2019) %V.Derya, S.Bagchi, J.Endres, M.N.Harakeh, J.Isaak, N.Kalantar-Nayestanaki, E.G.Lanza, B.Loher, A.Najafi, S.Pascu, S.G.Pickstone, N.Pietralla, V.Yu.Ponomarev, C.Rigollet, C.Romig, M.Spieker, A.Vitturi, A.Zilges
\end{thebibliography}

\end{document}